\documentclass[12pt]{article}
\usepackage[unicode=true,colorlinks=true,linkcolor=black, urlcolor=teal, citecolor = blue,breaklinks]{hyperref}
\usepackage{amsmath,amssymb,bm,bbm,amsthm}
\usepackage{graphicx}

\newcommand{\bra}[1]{\langle #1 |} 
\newcommand{\ket}[1]{| #1 \rangle}
\newcommand{\mi}{\imath}  
\newcommand{\me}{e} 
\newcommand{\e}{\bm{e}}
\newcommand{\Id}{\op{\mathbbm{1}}}
\newcommand{\Clf}{\bm{C\mspace{-2mu}l}} 
\newcommand{\Eq}[1]{Eq.~(\ref{#1})}
\newcommand{\Sec}[1]{Sec.~\ref{Sec:#1}} 
\newcommand{\Fig}[1]{Figure~\ref{Fig:#1}} 
\newcommand{\op}{\Hat}     
\newcommand{\Op}[1]{\,\widehat{\!#1\!}\,}     
\newcommand{\sgm}{\op{\sigma}}
\newcommand{\Ctrl}[1]{C\!#1}
\newcommand{\CX}{\Op{\Ctrl X}}
\newcommand{\CZ}{\Op{\Ctrl Z}}

\title{Mutual transformations of arbitrary ternary qubit trees by Clifford gates}
\date{}

\author{Alexander Yu.\ Vlasov}

\begin{document}

\maketitle

\begin{abstract}
 It is shown that ternary qubit trees with the same number of
 nodes can be transformed by the naturally defined sequence of 
 Clifford gates into each other or into standard representation 
 as one-dimensional chain corresponding to Jordan--Wigner transform.
\end{abstract}

\section{Introduction}
\label{Sec:Intro} 

The {\em ternary qubit tree} \cite{VlaQ3} is a visual representation of Clifford algebra
generators (or Majorana operators) as a multi-qubit Pauli operators.
It may be considered as a generalization of 
Jordan--Wigner transform \cite{JW} from a set of such operators naturally defined for 1D chain. 
The particular case of full ternary tree was suggested for optimal fermion-to-qubit mapping 
in quantum computations \cite{JKMN} and similar idea can be applied to more general ternary trees 
\cite{VQE,bonz}. 

Qubit trees provide a method to construct set of operators with necessary commutation 
properties, however for some applications it would be desirable also to provide a map
for conversion to standard Jordan--Wigner transform. The algorithm for representation 
of such map as a sequence of so-called stabilizer or Clifford gates \cite{Gott97,NCQC} is 
presented in this work. 

The definition of ternary qubit trees and relation with Jordan--Wigner transform
are recollected in \Sec{Clif}. The sequence of transformations of such
trees producing necessary stabilizer circuit is discussed in \Sec{Trans}.
Possible applications are briefly outlined in the last section.

\section{Clifford algebras and qubit trees}
\label{Sec:Clif}

Let us recollect definition of Clifford algebra \cite{ClDir} $\Clf(n)$ generated by 
$n$ elements with properties
\begin{equation}\label{ClAl}
\e_j \e_k + \e_k \e_j = 2 \delta_{jk}\Id,\quad j,k=1,\ldots,n,
\end{equation} 
where $\Id$ is unit of the algebra. Often right side of \Eq{ClAl} includes 
minus sign \cite{VlaQ3,ClDir}, but for complex Clifford algebra considered here such 
change is not essential due to possibility to multiply generators on imaginary unit.
The Clifford algebras defined here should not be confused with Clifford gates
also used in presented work.
The generators \Eq{ClAl} can be expressed using tensor product of Pauli matrices
\begin{equation}
\sgm^x = \begin{pmatrix}0&1\\1&0\end{pmatrix},\quad
\sgm^y = \begin{pmatrix}0&\!\!\!-\mi\\ \mi&\,0\end{pmatrix},\quad
\sgm^z = \begin{pmatrix}1&\,0\\0&\!\!\!-1\end{pmatrix}.
\label{PauliMat}
\end{equation}  
For even $n=2m$ such representation corresponds to Jordan--Wigner transform \cite{JW,ClDir}
\begin{equation*}
\begin{array}{rcl}
\e_{2k-1} & = &
{\underbrace{\sgm^z\otimes\cdots\otimes \sgm^z}_{k-1}\,}\otimes
\sgm^x\otimes\underbrace{\Id\otimes\cdots\otimes\Id}_{m-k} \\
\e_{2k} & = &
{\underbrace{\sgm^z\otimes\cdots\otimes \sgm^z}_{k-1}\,}\otimes
\sgm^y\otimes\underbrace{\Id\otimes\cdots\otimes\Id}_{m-k} 
\end{array},\qquad k=1,\ldots,m.
\end{equation*} 
It can be rewritten
\begin{equation}\label{ClGen}
\e_{2k-1} = \Bigl(\prod_{j=1}^{k-1}\sgm_j^z\Bigr)\sgm_k^x , \qquad
\e_{2k} = \Bigl(\prod_{j=1}^{k-1}\sgm_j^z\Bigr)\sgm_k^y,
\qquad k=1,\ldots,m.
\end{equation}
using formal notation
\begin{equation}
\label{sigk}
\sgm_k^\nu = {\underbrace{\Id\otimes\cdots\otimes \Id}_{k-1}\,}\otimes
\sgm^\nu\otimes\underbrace{\Id\otimes\cdots\otimes\Id}_{m-k}, \quad
k=1,\ldots,m,\quad\nu = x,y,z. 
\end{equation}
corresponding to action of Pauli matrix on a qubit with index $k$.

Due to \Eq{ClAl} the basis of Clifford algebra is $2^n$ different products of 
generators $\e_j$ and for $n=2m$ due to \Eq{ClGen} it coincides with $2^{2m}=4^m$ 
different tensor products of $m$ Pauli matrices. The \Eq{ClGen} is not the only
representation of elements with property \Eq{ClAl} as tensor product of Pauli matrices.

\begin{figure}[htb]
\begin{center}	
  \includegraphics[scale=0.75]{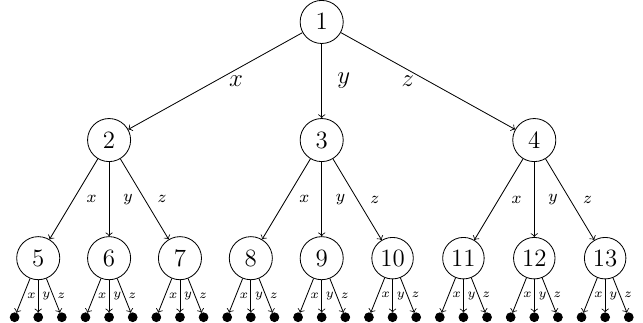}
\end{center}
\caption{Full ternary qubit tree}
\label{Fig:tern_tree}
\end{figure}

\begin{figure}[htb]
\begin{center}	
  \includegraphics[scale=0.75]{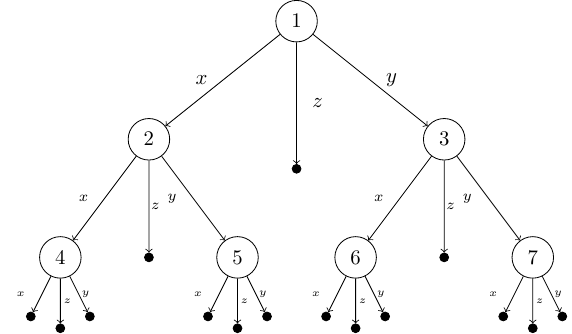}
\end{center}
\caption{Binary qubit tree augmented to ternary tree}
\label{Fig:bin_tree}
\end{figure}

Let us recollect the big class of alternative representations with so-called ternary qubit 
trees \cite{VlaQ3,JKMN,bonz}. Such trees have $m$ qubit nodes joined with three children by 
directed links marked by indexes $x,y,z$ and $2m+1$ terminal nodes. An example of full ternary tree
of such kind is represented on \Fig{tern_tree}, but arbitrary tree with up to three children 
nodes could be augmented to such ternary tree by attaching missing terminal nodes. An example of
such ternary tree constructed from a binary one is shown on \Fig{bin_tree}.

Let us now associate $\sgm_k^\nu$ with a link from qubit $k$ with a label $\nu$.
Then each path $P$ from a root to a terminal node corresponds to product of 
single-qubit Pauli operators \Eq{sigk} along this 
path according to the indices of the qubit nodes and the labels assigned to
the outgoing arrows 
\begin{equation}
\label{e33}
\e_P = \prod_{P} \sgm_{k(P)}^{\nu(P)}.
\end{equation}
Any two such products $\e_P$ and $\e_{P'}$ include only one Pauli matrix with the same $k$
and different $\nu\neq\nu'$. Here $k$ is an index of last qubit in common part of paths $P$ and $P'$ 
before forking. Thus, any two unequal products \Eq{e33} anticommute and satisfy `nontrivial
part' $\e_j\e_k=-\e_k\e_j$, $j \neq k$ of \Eq{ClAl}. The `trivial part' $\e_j^2=\Id$ is 
also obviously satisfied.

For example, the full ternary tree on \Fig{tern_tree} corresponds to 27 products
$$\e_1 = \sgm^x_1\sgm^x_2\sgm^x_5, 
\ \e_2 = \sgm^x_1\sgm^x_2\sgm^y_5, 
\ \e_3 = \sgm^x_1\sgm^x_2\sgm^z_5,
\ \e_4 = \sgm^x_1\sgm^y_2\sgm^x_6, \ldots,
\ \e_{27} = \sgm^z_1\sgm^z_4\sgm^z_{13}
$$ 
and binary tree on \Fig{bin_tree} produces 15 generators 
$$\e_1 = \sgm^x_1\sgm^x_2\sgm^x_4, 
\ \e_2 = \sgm^x_1\sgm^x_2\sgm^z_4, 
\ \e_3 = \sgm^x_1\sgm^x_2\sgm^y_4,
\ \e_4 = \sgm^x_1\sgm^z_2, \ldots,
\ \e_{15} = \sgm^y_1\sgm^y_3\sgm^y_7.
$$

There are $2m+1$ such products corresponding to number of terminal nodes, but arbitrary one 
can be dropped, because up to unessential ``phase'' any such $\e_P$ can be represented as 
product of other $2m$ terms \cite{VlaQ3}.
For usual Jordan--Wigner transform \Eq{ClGen} the extra term would correspond to  
\begin{equation}\label{ClGenProd}
\e_{2m+1} = \prod_{j=1}^{k}\sgm_j^z.
\end{equation}

\section{Transformations between representations}
\label{Sec:Trans}

For any unitary operator $\op U$ {\em adjoint action} ({\em conjugation} by $\op U$) produces new 
set of generators \begin{equation}
\label{TranGen}
 \mathrm{Ad}_{\op U} \colon \e_j \mapsto \op U \e_j \op U^{-1} = \op U \e_j \op U^\dag,\quad j=1,\ldots,n
\end{equation}
satisfying necessary commutation relations \Eq{ClAl}. For $\op U$ represented by
so-called {\em quantum stabiliser circuits} composed from product of Clifford 
gates \cite{Gott97,NCQC} \Eq{TranGen} represents transformation 
between $4^k$ different products of $\sgm_k^\nu$ \Eq{sigk}. Such a property was 
discussed earlier for application to construction of new quantum gates effectively
modelled by classical computer \cite{JM8}.

Let us recollect construction of Clifford (stabilizer) gates \cite{Gott97,NCQC}. 
For a single qubit the group transforming $\sgm^\nu$ into each other by conjugation 
may be generated by Pauli matrices $\sgm^\nu$ together with Hadamard matrix $\op H$ and
`phase gate' $\op S$
\begin{equation}
\label{HS}
 \op H = \begin{pmatrix}1 & \,\,1 \\ 1 & \!\!\!-1 \end{pmatrix}, 
 \quad \op S = \begin{pmatrix}1 & 0 \\ 0 & \mi\end{pmatrix}.
\end{equation}
Conjugation by Pauli matrices can only change a sign and neglected further, 
but $\op H$ and $\op S$ exchanges $\sgm^x \leftrightarrow\sgm^z$
and $\sgm^x \leftrightarrow\sgm^y$ respectively.

For two qubits together with  
$\op H_j$ and $\op S_j$, $j=1,2$ acting on any of them it is necessary to include two-qubit gate 
Controlled-NOT (${\CX}$) or Controlled-$Z$ (${\CZ}$) to generate whole group of stabilizer 
transformations
\begin{equation}
\label{CXZ}
 {\CX} = \begin{pmatrix}1 & 0 & 0 & 0 \\ 0 & 1 & 0 & 0 \\
                     0 & 0 & 0 & 1 \\ 0 & 0 & 1 & 0 \end{pmatrix}, \quad 
 {\CZ}= \begin{pmatrix}1 & 0 & 0 & \,\,0 \\ 0 & 1 & 0 & \,\,0 \\
                     0 & 0 & 1 & \,\,0 \\ 0 & 0 & 0 & \!\!\!-1 \end{pmatrix}.
\end{equation}
For arbitrary number of qubits compositions of gates $\op H_j$, $\op S_j$ acting on arbitrary
qubit $j$ together with ${\CX}_{jk}$ or ${\CZ}_{jk}$ acting on arbitrary pair of qubits $(j, k)$
generate whole group of transformations preserving products $\sgm_k^\nu$ (and
basis of Clifford algebra). 

The exchange between arbitrary pair of qubits also may be
decomposed on such operators, but for convenience it can be included as a separate gate.
The operator ${\CZ}_{jk}$ is used further due to convenience for some transformations of 
qubit trees, symmetry of two indexes and commutativity for different qubits. 

Conjugation by ${\CZ}_{12}$ for all possible tensor products of Pauli (and 
identity) matrices are collected in the table below
\begin{equation}
\label{CZmap}
{\setlength{\arraycolsep}{0.2em} 
\!\!\begin{array}{|l|l|l|l|}\hline
\vphantom{\bigm|^|} \Id \otimes \Id\ \rightarrow \Id \otimes \Id &
\sgm^x \otimes \Id \rightarrow \sgm^x {\otimes} \sgm^z &
\sgm^y \otimes \Id \rightarrow \sgm^y {\otimes} \sgm^z &
\sgm^z \otimes \Id \rightarrow \sgm^z \otimes \Id \\
\Id \otimes \sgm^x \rightarrow \sgm^z {\otimes} \sgm^x &
\sgm^x {\otimes} \sgm^x \rightarrow \sgm^y {\otimes} \sgm^y &
\sgm^y {\otimes} \sgm^x \rightarrow \sgm^x {\otimes} \sgm^y &
\sgm^z {\otimes} \sgm^x \rightarrow \Id \otimes \sgm^x\\ 
\Id \otimes \sgm^y \rightarrow \sgm^z {\otimes} \sgm^y &
\sgm^x {\otimes} \sgm^y \rightarrow \sgm^y {\otimes} \sgm^x &
\sgm^y {\otimes} \sgm^y \rightarrow \sgm^x {\otimes} \sgm^x &
\sgm^z {\otimes} \sgm^y \rightarrow \Id \otimes \sgm^y\\ 
\Id \otimes \sgm^z \rightarrow \Id \otimes \sgm^z &
\sgm^x {\otimes} \sgm^z \rightarrow \sgm^x {\otimes} \Id &
\sgm^y {\otimes} \sgm^z \rightarrow \sgm^y {\otimes} \Id &
\sgm^z {\otimes} \sgm^z \rightarrow \sgm^z {\otimes} \sgm^z\\
\hline
\end{array}\!\!
}
\end{equation}

Let us consider action of stabilizer gates on the qubit trees.
The one-qubit gates such as $\op H_j$, $\op S_j$ and products of such operators
map one qubit tree into another and simply exchange labels $x,y,z$ of links 
attached to qubit $j$. For system with two qubits any set of generators
may be represented as ternary tree and so two qubit-gates ${\CZ}$
also corresponds to some transformation of the trees. 

For three and more qubits two-qubit gates such as ${\CZ}_{jk}$ may 
transform products representing $\e_j$ for some ternary qubit tree into collection
that either represent another tree or not. For example let us consider
standard six generators \Eq{ClAl} for three qubits
\begin{equation}
\label{ClGen6}
 \e_1 = \sgm^x_1,\  \e_2 = \sgm^y_1,
 \  \e_3 = \sgm^z_1\sgm^x_2,\  \e_4 = \sgm^z_1\sgm^y_2,
 \  \e_5 = \sgm^z_1\sgm^z_2\sgm^x_3,\  \e_6 = \sgm^z_1\sgm^z_2\sgm^y_3.
\end{equation}
Conjugation by $\op U = {\CZ}_{1\,2}\,{\CZ}_{2\,3}$ produces six generators
mentioned in Ref.~\cite{bonz}
with correct commutation relation \Eq{ClAl}, but without representation
as a ternary qubit tree
\begin{equation}
\label{ClGen6ex}
 \tilde\e_1 = \sgm^x_1\sgm^z_2,\  \tilde\e_2 = \sgm^y_1\sgm^z_2,
 \  \tilde\e_3 = \sgm^x_2\sgm^z_3,\  \tilde\e_4 = \sgm^y_2\sgm^z_3,
 \  \tilde\e_5 = \sgm^z_1\sgm^x_3,\  \tilde\e_6 = \sgm^z_1\sgm^y_3.
\end{equation}

Anyway, ${\CZ}_{jk}$ also may produce valid transformation of ternary tree
and it is shown further, how to map any ternary tree
$T$ into standard (Jordan--Wigner) form corresponding to \Eq{ClGen} using 
\Eq{TranGen} with some operator denoted here as $\op U_T^\dag$ with the inverse 
transformation $\op U_T$ mapping \Eq{ClGen} into set of generators for $T$. 
Arbitrary tree $T$ also can be converted into another 
$T'$ by composition of two such transformations $\op U_T^\dag \op U_{T'}$ with possible 
further optimization to reduce number of gates.

\begin{figure}[htb]
\begin{center}
\parbox{0.45\textwidth}{$a$)\includegraphics[scale=0.35]{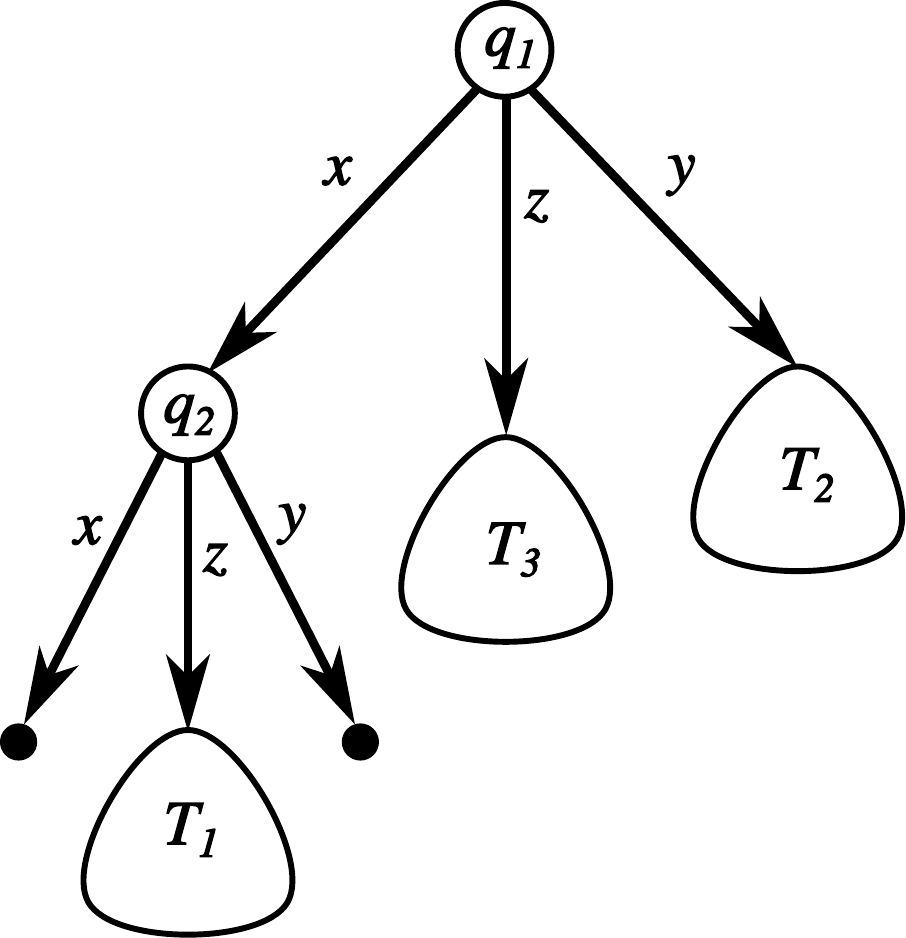}}%
$\longrightarrow $%
\parbox{0.45\textwidth}{$b$)\includegraphics[scale=0.35]{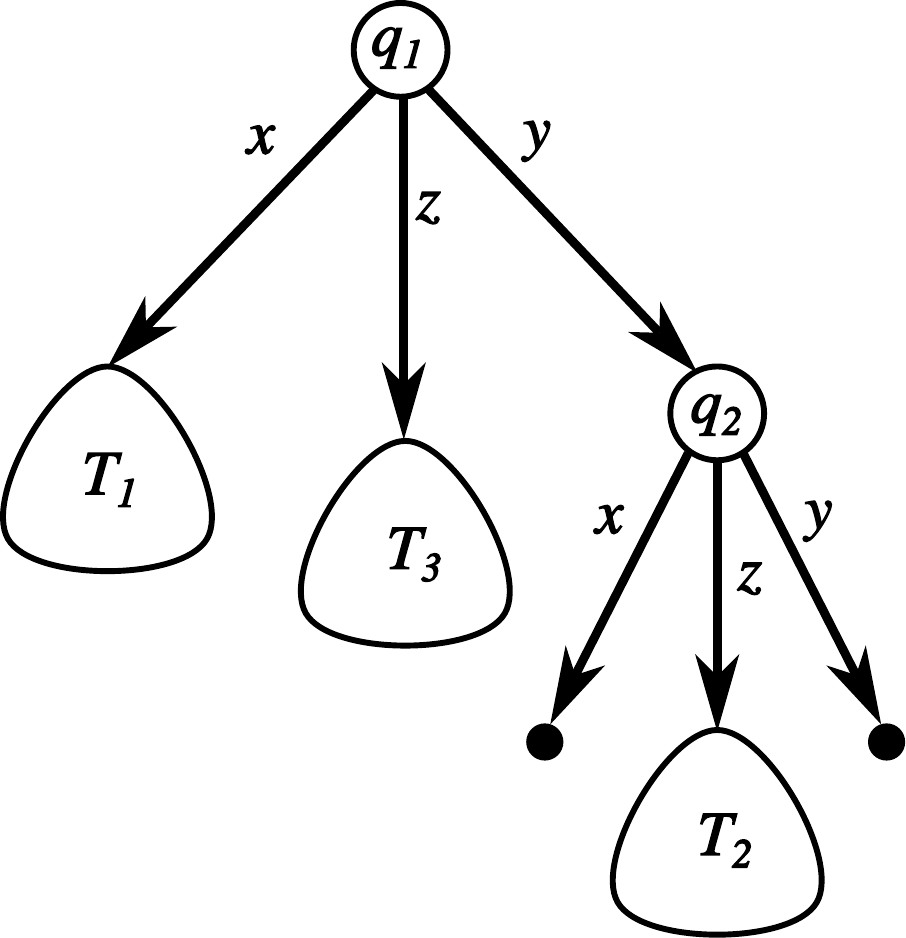}}
\end{center}
\caption{Transformation of qubit tree by $\CZ_{q_1 q_2}$}
\label{Fig:trans_tree}
\end{figure}

The idea is to `straighten all forks' in the tree. Let us consider
the last fork on one of the branches originated from some qubit
with index $q_1$. It is depicted on \Fig{trans_tree}$a$.
It is supposed, that there are no more forks on the branches
attached to the qubit $q_1$, but this branches can be either terminal
or include arbitrary number of qubit nodes. For certainty the 
branch attached by link with label $x$ has amount 
of nodes less or equal than others.

Let us show that conjugation by gate ${\CZ}_{q_1q_2}$ rejoins the closest qubit 
node with index $q_1$ to the link with the label $y,$ see \Fig{trans_tree}$b$. 
Here we suppose that node $q_2$ has two
terminal nodes attached by links $x$ and $y$ with some tree $T_1$ may be 
attached to link $z$. It is always possible to exchange labels 
in such a way by one-qubit transformation. Some trees $T_2$ and
$T_3$ also may be attached to qubit node $q_1$ by links
with labels $y$ and $z$.

The scheme of tensor products of Pauli (and identity) matrices for generators corresponding
to ternary qubit tree on \Fig{trans_tree}$a$ is presented on the left-hand side
of \Eq{forks}.
\begin{equation}
\label{forks}
\begin{array}{l}
\cdots\,\sgm^x \otimes \sgm^x \otimes \Id\ \cdots\, \cdots \\
\cdots\,\sgm^x \otimes \sgm^y \otimes \Id\ \cdots\, \cdots \\
\cdots\,\sgm^x \otimes \sgm^z \otimes T_1 \cdots\, \cdots \\
\cdots\,\sgm^y \otimes \,\Id\ \otimes \cdots T_2 \cdots\\
\cdots\,\sgm^z \otimes \,\Id\ \otimes \cdots \cdots T_3 \\
\end{array}
\quad\longrightarrow\quad
\begin{array}{l}
\cdots\,\sgm^y \otimes \sgm^y \otimes \Id\ \cdots\, \cdots \\
\cdots\,\sgm^y \otimes \sgm^x \otimes \Id\ \cdots\, \cdots \\
\cdots\,\sgm^x \otimes \,\Id\ \otimes T_1 \cdots\, \cdots \\
\cdots\,\sgm^y \otimes \sgm^z \otimes \cdots T_2 \cdots\\
\cdots\,\sgm^z \otimes \,\Id\ \otimes \cdots \cdots T_3 \\
\end{array}
\end{equation}
Here non-overlapping commuting terms $T_1$, $T_2$, $T_3$ correspond
to subtrees on \Fig{trans_tree}. They could be arbitrary, yet only
subtrees without forks are enough further. The positions corresponding to 
qubits $q_1$ and $q_2$ for simplicity are arranged sequentially
in \Eq{forks} and corresponds to combinations
\[
\sgm^x \otimes \sgm^x,\quad \sgm^x \otimes \sgm^y,%
\quad \sgm^x \otimes \sgm^z,\,\quad \sgm^y \otimes \Id,%
\quad \sgm^z \otimes \Id
\]
mapping due to \Eq{CZmap} into
\[
\sgm^y \otimes \sgm^y,\quad \sgm^y \otimes \sgm^x,%
\,\quad \sgm^x \otimes \Id,\quad \sgm^y \otimes \sgm^z,%
\quad \sgm^z \otimes \Id
\]
respectively. The result of such map is shown on the right-hand side of \Eq{forks}
and corresponds to qubit tree on \Fig{trans_tree}$b$.

\begin{figure}[htbp]
\begin{center}	
  \includegraphics[width=\textwidth]{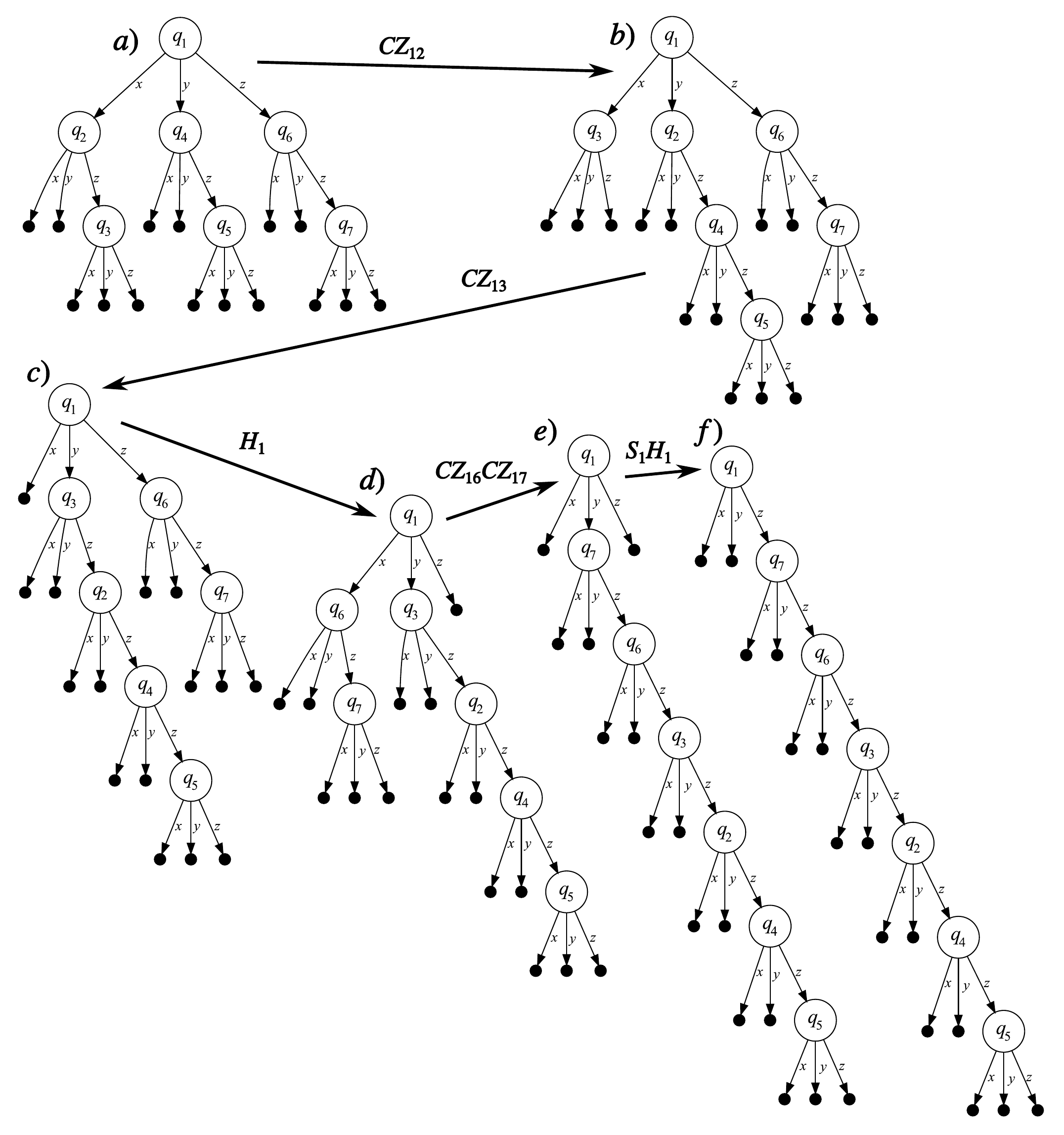}
\end{center}
\caption{Example of ternary qubit trees transformations by Clifford gates}
\label{Fig:trees7x6}
\end{figure}

Example of conversions from a triple fork to a linear chain is shown on \Fig{trees7x6}. 
Three branches with two nodes are attached to qubit $q_1$ by $x$, $y$ and $z$ links on 
\Fig{trees7x6}$a$. 
The gate ${\CZ}_{1\,2}$ is used to move one qubit node from $x$ to $y$ branch,  
see \Fig{trees7x6}$b$. The gate ${\CZ}_{1\,3}$ moves yet another qubit node from $x$ to $y$ and on 
\Fig{trees7x6}$c$ now the $x$ branch has a single terminal node. To continue process it
is necessary to exchange branches $x$ and $z$ by gate $\op H_1$, see \Fig{trees7x6}$d$.
Now it is possible to move qubit nodes $q_6$ and $q_7$ from $x$ to $y$ branch
using gates ${\CZ}_{1\,6}$ and ${\CZ}_{1\,7}$,  see \Fig{trees7x6}$e$.
Now only $y$-branch has qubit nodes and gates $\op S_1$ and $\op H_1$ are used
finally to exchange $y$ and $z$ links to produce a chain of $z$-linked nodes, 
see \Fig{trees7x6}$f$.

Consequent application of similar operations to all forks on a ternary qubit tree produces
chain without any branches with nodes connected by $z$-links. It is only necessary
to rearrange qubit indexes in ascending order to produce `degenerate' qubit tree
corresponding to a chain and standard (Jordan--Wigner) representation \Eq{ClGen}
of generators $\e_j$.

\section{Conclusion}
\label{Sec:Concl}

Adjoint map \Eq{TranGen} between standard Jordan--Wigner transform \Eq{ClGen} and 
representation by arbitrary ternary qubit tree \Eq{e33} using the stabilizer circuit
(the sequence of Clifford gates) is constructed 
in this work. The applications of such map can be numerous. A general idea
of a few such applications is similar with mentioned in Ref.~\cite{JM8} for 
some other examples of quantum circuits `intertwined by Clifford operations'.
It was initially suggested for effective simulations of specific nonuniversal quantum 
circuits by classical computer, but some ideas recollected below can be useful
for other applications.

The approach with conjugation by Clifford operations also looks more general, 
because not all set of generators expressed as \Eq{TranGen} can be represented 
by ternary trees.
However, such ideas appropriate for any set of generators represented
using conjugation \Eq{TranGen} and so may be applied to ternary trees as well
due to construction discussed in \Sec{Trans}.

Adjoint transformation $\op G'=\mathrm{Ad}_{\op U}\op G$ can be associated 
with transition from a quantum circuit $\op G$ acting on a state $\ket{\Psi}$ to a new circuit
$\op G'$ for a quantum state $\ket{\Psi'} = \op U \ket{\Psi}$ with trivial substitution
\[\op G'\ket{\Psi'} = \op U\op G \op U^\dag\ket{\Psi'} =  \op U\op G\op U^\dag\,\op U \ket{\Psi} 
= \op U \op G \ket{\Psi}. \] 

Any observable $\op O$ used
in the measurement $\bra{\Psi}\op O\ket{\Psi}$ for some states in initial circuit 
would correspond to new observable $\op O' = \op U \op O \op U^\dag$. 
If such observables was expressed  
as an analytical function of generators $\e_k$, such as quadratic, polynomial
or exponential, the new observables also can be expressed in the same way.
Similar property also valid for Hamiltonians $H$ used for formal expression
of quantum evolution 
\[\ket{\Psi(t)} = \op G(t)\ket{\Psi_0} = \me^{-\mi \op H t}\ket{\Psi_0},\] 
where for modified quantum system or circuit a Hamiltonian can be 
expressed as $\op H' = \op U \op H \op U^\dag$. 
Here the general ideas of possible applications are only outlined
and concrete examples should be discussed elsewhere.

\end{document}